\documentclass[aps,nofootinbib,amsmath,prd,twocolumn,showpacs,superscriptaddress,groupedaddress]{revtex4-1}

\let\oldtitle\title
\renewcommand{\title}[1]{\oldtitle{\color{blue}{#1}}}

\usepackage{amsfonts}
\usepackage{amsmath}
\usepackage{amssymb}
\usepackage{epsfig}
\usepackage{latexsym}
\usepackage{paralist}
\usepackage{fancyhdr}
\usepackage{graphicx}
\usepackage[vcentermath]{youngtab}
\usepackage{young}
\usepackage{ytableau}
\usepackage{etex}
\usepackage{braket}
\usepackage{float}
\usepackage{slashed}
\usepackage{xcolor}
\usepackage{soul}
\usepackage{dcolumn} 
\usepackage{physics}

\def\bea{\begin{eqnarray}} 
\def\eea{\end{eqnarray}}
\def\be{\begin{equation}} 
\def\ee{\end{equation}} 
\def\ba{\begin{array}}
\def\ea{\end{array}}

\def\be{\begin{equation}}
\def\ee{\end{equation}}
\def\bea{\begin{eqnarray}}
\def\eea{\end{eqnarray}}

\usepackage{amsmath}

\begin{document}

\title{Renormalization of multicritical scalar models in curved space}

\author{Riccardo Martini}
\email{riccardo.martini@uni-jena.de}
\affiliation{
Theoretisch-Physikalisches Institut, Friedrich-Schiller-Universit\"{a}t Jena,
Max-Wien-Platz 1, 07743 Jena, Germany}

\author{Omar Zanusso}
\email{omar.zanusso@uni-jena.de}
\affiliation{
Theoretisch-Physikalisches Institut, Friedrich-Schiller-Universit\"{a}t Jena,
Max-Wien-Platz 1, 07743 Jena, Germany}

\begin{abstract}
We consider the leading order perturbative renormalization of the multicritical $\phi^{2n}$ models and some generalizations in curved space.
We pay particular attention to the nonminimal interaction with the scalar curvature $\frac{1}{2}\xi \phi^2 R$
and discuss the emergence of the conformal value of the coupling $\xi$
as the renormalization group fixed point of its beta function
at and below the upper critical dimension as a function of $n$.
We also examine our results in relation with Kawai and Ninomiya's formulation of two dimensional gravity.
\end{abstract}

\maketitle

\renewcommand{\thefootnote}{\arabic{footnote}}
\setcounter{footnote}{0}

\section{Introduction}\label{sect:introduction}

The multicritical scalar models with $\phi^{2n}$ interaction are the simplest
and most straightforward generalization of the $\phi^4$ model.
Much like the $\phi^4$ field theory captures the critical properties of an universality class
of models that includes the ferromagnetic Ising Hamiltonian,
the $\phi^{2n}$ field theory can be thought as describing a generalization
in which the Ising's spin domains of plus or minus sign are potentially replaced
by $n$ distinct vacuum states which become degenerate
at the critical temperature.

The renormalization group (RG) flow of the $\phi^{2n}$ models
has been explored at length in the literature:
perturbatively \cite{Itzykson:1989sx,ODwyer:2007brp,Codello:2017hhh}, nonperturbatively \cite{Morris:1994jc,Codello:2012sc,Hellwig:2015woa},
and with non-canonical kitetic terms \cite{Safari:2017irw,Safari:2017tgs}.
It is well-known that a consistent perturbative expansion in the coupling
can be constructed at the upper critical dimension
\begin{equation}\label{eq:upper-critical-dimension}
\begin{split}
 d_n= \frac{2n}{n-1}\,.
\end{split}
\end{equation}
It is easy to check that, as expected, the case $n=2$ corresponds to the $\phi^4$
interaction which has upper critical dimension $d=4$ \cite{Wilson:1973jj}. The model $n=3$ corresponds
to the $\phi^6$ interaction and is known to describe the universal features of the
tricritical Ising model with upper critical dimension $d=3$ \cite{Hager:2002uq}.
All other models have purely fractional upper critical dimensions \cite{Gracey:2017okb} which asymptotically
tend to $d=2$. As a consequence $d=2$
is the first physical dimension in which \emph{all} the models $\phi^{2n}$
are nontrivial; the continuation to two dimensions
is particularly relevant because they are known to interpolate with the unitary minimal models
arising as representations of the infinite dimensional Virasoro algebra \cite{Belavin:1984vu,Zamolodchikov:1987ti}.

For the most part the renormalization of the multicritical models
generalizes the one of the $\phi^4$ model, but the leading contributions
to the critical exponents are determined by multiloop computations
in which the number of loops increases with $n$ \cite{ODwyer:2007brp}.
Likewise the $\phi^4$ model, the $\phi^{2n}$ interactions describe critical theories
that are Gaussian for $d>d_n$ and logarithmic at $d=d_n$, but have non-trivial critical exponents for $d<d_n$.
A common practice is to compute such critical exponents in the $\epsilon$-expansion,
in which one introduces the constant $\epsilon=d_n-d$ and uses it to parametrize the displacement of the critical point
from the Gaussian theory
at $d=d_n$ \cite{Wilson:1973jj}.

The only multicritical model that has nontrivial exponents in $d=3$ is the $\phi^4$ one
unless one includes the multicritical non-unitary models $\phi^{2n+1}$ \cite{Codello:2017epp}, as we shall briefly do later.
Specifically, $\phi^{3}$ and $\phi^{5}$ have upper critical dimensions $d=6$ \cite{Fisher:1978pf} and $d=\frac{10}{3}$ respectively,
but they require the tuning of an imaginary-valued magnetic field at criticality \cite{vonGehlen:1989yn,vonGehlen-2,Zambelli:2016cbw}.
It is important to mention that
the $d=2$ realizations of these models are all ``far away'' in a perturbative sense
from their Gaussian points even though $d_{n}\to 2$ for $n\to\infty$ \cite{ODwyer:2007brp}.
Nevertheless, the simple existence of the sequence of multicritical theories
provides a very interesting and valuable link between
purely field-theoretical realizations and CFT representations \cite{Codello:2017qek,Codello:2017hhh}.

One natural and potentially interesting generalization of the above discussion
is the study of the renormalization of the $\phi^{2n}$ models in \emph{curved space}.
Generically, the renormalization of a model in a curved background requires additional care to preserve covariance
and further conditions to avoid new and unwanted infinities \cite{Brown:1980qq}.
The extra work is often a gateway to extra information on the theory under consideration \cite{Jack:1983sk}.
If the multicritical models are coupled with a background geometry, simple dimensional analysis reveals that
there is a new non-minimal marginal interaction with the curvature: $\frac{1}{2}\xi \phi^2 R$.
One expects that in curved space the perturbative construction should thus accommodate for some mixing
between the $\phi^{2n}$ and $\phi^2 R$ operators regardless of $n$.
In other words, the nonminimal interaction $\phi^2 R$ holds a special status
in that it is always canonically marginal.

A guess on the value that the coupling $\xi$ can take at a curved space generalization
of the critical point could be made as follows:
Consider a nonminimally coupled ``free'' scalar field with quadratic action
\begin{equation}\label{eq:free-nonminimal-action}
\begin{split}
 S_0[\phi] &= \frac{1}{2}\int {\rm d}^dx \Bigl\{ g^{\mu\nu}\partial_\mu\phi\partial_\nu\phi + \xi \phi^2 R\Bigr\}\,.
\end{split}
\end{equation}
Ideally, the above action captures the Gaussian (non-self-interacting) limit of the $\phi^{2n}$ models
which is realized exactly at the upper critical dimension.
The nonminimal action is invariant under a conformal Weyl rescaling $g'_{\mu\nu}= \Omega^2(x) g_{\mu\nu}$ and $\phi'(x)=\Omega^{1-\frac{d}{2}}(x) \phi(x)$
iff the coupling $\xi$ takes the \emph{conformal value}
\begin{equation}\label{eq:xi-conformal}
\begin{split}
 \xi_{c} &= \frac{d-2}{4(d-1)}\,.
\end{split}
\end{equation}
Since conformal invariance implies scale invariance, the nonminimal action \eqref{eq:free-nonminimal-action}
is thus scale invariant when $\xi$ takes the conformal value, but it
is also expected to be a description of the critical (scale-invariant) $\phi^{2n}$ model
when the interaction becomes Gaussian at the upper critical dimension.
Putting everything together we make the following guess.

\begin{quote}
{\bf Educated guess:}
The critical point of the coupling $\xi$ at the upper critical dimension,
which emerges as fixed point of the renormalization group, is the conformal
value \eqref{eq:xi-conformal}.%
\footnote{
The conformal invariance of \eqref{eq:free-nonminimal-action} is actually expected to be anomalous \cite{Mottola:1995sj},
but for our purposes it is sufficient that scale invariance survives the quantization process.
In even dimensions, the anomaly is signaled by special nonlocal contributions
appearing in the effective action \cite{Distler:1988jt,Ribeiro:2018pyo}.
}
\end{quote}

More generally, one would be tempted to extend the above statement to any dimension
\emph{below} the upper critical dimension, having expressed the desire of analytically continuing these models to $d=2$.
In this case, we would want to know the conditions under which the conformal value \eqref{eq:xi-conformal} is always the critical
value for $\xi$ even when the $\phi^{2n}$ interaction is non-Gaussian below $d_c$.
For this purpose it is instructive to recall the investigation by Brown and Collins \cite{Brown:1980qq},
in which it is shown that at the leading order our educated guess is true in the special case of the $\phi^4$ model,
but beyond the leading order one has to exploit the freedom of subtracting additional finite parts
proportional to the leading counter terms \cite{Jack:1983sk}.
An analog renormalization condition has also been adopted for the $\phi^3$ model \cite{Jack:1985wf} in $d=6$,
and it plays an important role in preserving conformal invariance in \cite{Grinstein:2014xba,Grinstein:2015ina}.
Notice that, strictly speaking, the said two examples have not been concerned with the analytic continuation
below the upper critical dimensions $d=4$ and $d=6$, while our interest is to bring the multicritical models
down to $d=2$ which does require continuation.
However, assuming that we have the same freedom in changing the renormalization condition, we can state a conjecture.


\begin{quote}
{\bf General conjecture:}
The critical point of the coupling $\xi$, which perturbatively is determined
as an $\epsilon$-expansion in $\epsilon=d_n-d$, can always be the conformal
value \eqref{eq:xi-conformal} thanks to an opportune renormalization condition.
\end{quote}

In this paper we consider the leading renormalization and $\epsilon$-expansion
of all the infinitely many multicritical models $\phi^{2n}$
(and some other generalizations as well) in curved space
using the formalism of functional perturbation theory \cite{ODwyer:2007brp,Codello:2017hhh}.
With the leading results we can show that the conformal value \eqref{eq:xi-conformal}
of $\xi$ is indeed the critical value at the leading order in $\epsilon$,
thus proving the educated guess of this introduction.
While we leave the above general conjecture open, we stress that the educated guess
is proven for an infinite number of theories.
Interestingly, our computation is genuinely new in that the structure of
the counterterms and their renormalization does not come from a straightforward
generalization of the $\phi^4$ case.
On the contrary, we see the case $n=2$ as quite the exception which we have to deal with separately.

The paper is organized as follows:
In Sect.~\ref{sect:renormalization} we study divergences, counterterms, and renormalization group beta functions
for all the $\phi^{2n}$ models. We discuss separately the cases $n=2$, $n=\infty$ and the nonunitary models $\phi^{2n+1}$.
We elaborate briefly on the utility of our results in reproducing some well known formula of $2d$ gravity in the limit of large central charge.
In Sect.~\ref{sect:criticality} we show how the conformal value of $\xi$ emerges as fixed point of its beta function.
We also elaborate more on the stronger conjecture expressed in this introduction.
Finally in Sect.~\ref{sect:conclusions} we draw some conclusion and give a prospect for future investigations.
The appendices are dedicated to technical details on the covariant renormalization in curved space.
In particular, appendix~\ref{sect:heat-kernel} discusses the Seeley-de Witt representation of the covariant Green function,
and appendix~\ref{sect:feynman} briefly describes an algorithm by Jack and Osborn for the computation of the poles of dimensionally regulated
covariant Feynman diagrams in curved space.

\section{Renormalization}\label{sect:renormalization}

We are interested in a simple self-interacting canonically normalized scalar field $\phi$
which is nonminimally coupled to a background metric $g_{\mu\nu}$ in $d$ dimensions.
The straightforward bare action is
\begin{equation}\label{eq:bare-action}
\begin{split}
 S[\phi] = \int {\rm d}^dx \sqrt{g}\Bigl\{
 \frac{1}{2}g^{\mu\nu}\partial_\mu\phi\partial_\nu\phi +V(\phi) +F(\phi) R
 \Bigr\}\,.
\end{split}
\end{equation}
Using the bare action we can formally construct the path integral.
For later convenience we shall do it in the background field approach, thus by integrating
the fluctuations $\chi$ over an arbitrary background $\phi$ as follows
\begin{equation}
\begin{split}
 Z = \int {\rm D}\chi ~ {\rm e}^{-S[\phi+\chi]}\,.
\end{split}
\end{equation}

In flat space it is possible to construct a meaningful perturbative expansion
for potentials $V(\phi)$ which are polynomials of order $2n$ below the upper critical dimensions \eqref{eq:upper-critical-dimension}.
If we parametrize $V(\phi)=\frac{\lambda}{(2n)!}\phi^{2n} + \dots$, the upper critical dimension
is the one for which the canonical dimension of $\lambda$ is zero, and the perturbative expansion
is controlled by powers of $\lambda$ itself. Below the upper critical dimensions,
these perturbative expansions are known to lead to a sequence of universality classes
often referred to as \emph{minimal models} because they interpolate with the minimal conformal theories
arising as representations of the Virasoro algebra in $d=2$ \cite{Belavin:1984vu}.

A simple dimensional analysis reveals that if we parametrize $F(\phi)=\frac{\xi}{2}\phi^2 +\dots$,
the coupling $\xi$ is \emph{always} dimensionless and thus it is expected to play a role in the perturbative
expansion when promoting the minimal models to curved space. In other words, the $\phi^{2n}$ and $\phi^2 R$ operators are
expected to mix because of statistical or quantum mechanical fluctuations.
For the above reasons we are interested in renormalizing the path integral in $d=d_n$ dimensions
with $V(\phi)$ and $F(\phi)$ restricted to be polynomials of order $2n$ and $2$ respectively,
so to include all the relevant and naively marginal operators of the models. We do it by adopting dimensional
regularization which corresponds to analytically continuing the dimensionality to $d=d_n-\epsilon$.

Since the order of the nonminimal interaction is only two, we can incorporate it easily in a quadratic part of the
bare action
\begin{equation}
\begin{split}
 S_0[\chi] = \frac{1}{2}\int {\rm d}^dx \sqrt{g}
 \chi\left(-\nabla^\mu \partial_\mu +F''(\phi) R\right)\chi
 \,.
\end{split}
\end{equation}
According to the dimensionality, there are two possible leading contributions if the action of the path integral is expanded perturbatively around $S_0[\chi]$
for the $\phi^{2n}$ models in powers of $V(\phi)$: the linear and the quadratic contributions.

Expanding the path integral to the linear order in $V(\phi+\chi)$ and Taylor-expanding
the potential itself we have a generalized tadpole-like contribution
\begin{equation}\label{eq:linear-path-integral-contribution}
\begin{split}
 - \int {\rm d}^dx \sqrt{g(x)} ~ \sum_{0\leq r \leq n}
 \frac{1}{(2r)!} ~ G(x,x)^r ~ V^{(2r)}(\phi(x))
 \,,
\end{split}
\end{equation}
in which the number of closed lines is constrained to be even because of trivial topological reasons.
In dimensional regularization the linear term contributes to the renormalization of the potential
only if $r=1$ and $d=2$ as we show later in subsection \ref{subsect:sg}.
At the quadratic order we have instead
\begin{equation}\label{eq:leading-path-integral-contribution}
\begin{split}
 &\frac{1}{2} \int {\rm d}^dx ~ {\rm d}^dx' \sqrt{g(x)g(x')} ~ \times \\
 &\times \sum_{0\leq r \leq 2n}
 \frac{1}{r!} V^{(r)}(\phi(x)) ~ G(x,x')^r ~ V^{(r)}(\phi(x'))
 \,.
\end{split}
\end{equation}

In \eqref{eq:linear-path-integral-contribution} and \eqref{eq:leading-path-integral-contribution} we introduced $G(x,x')$
which is the Green function associated to the operator of the quadratic part of the action
\begin{equation}
\begin{split}
 &{\cal O} = - g^{\mu\nu}\nabla_\mu\partial_\nu + F''(\phi) R\,, \\
 &{\cal O}_x G(x,x') = \delta^{(d)}(x,x')\,.
\end{split}
\end{equation}
A covariant representation of the Green function for an operator of Laplace-type as the one above is described in Appendix \ref{sect:heat-kernel}.
For our present needs, the representation simply shows that the Green function can be expanded
\begin{eqnarray}
 G(x,x') &=& G_0(x,x') + a_1(x,x') ~ G_1(x,x')
 +\dots \,,
\end{eqnarray}
in which we purposely neglected all further contributions which do not affect the relevant operators.
The leading $G_0(x,x')$ term can be understood as a covariant generalization of the standard
Green function of flat space (see Appendix \ref{sect:heat-kernel} for more details),
while $a_1(x,x')$ is the first correction due to curvatures and multiplies the subleading correction to the propagator $G_1(x,x')$.

In the following subsection we consider first the renormalization of the general $\phi^{2n}$ universality class for $n\geq 3$,
while the case $n=2$ is deferred for later. The reason for this is that the case $n=2$ is special when it comes to the
renormalization of the function $F(\phi)$. In particular, the results for the general $\phi^{2n}$ case often cannot be continued to $n=2$
because the subleading correction to the propagator is powerlaw for each $d=d_n$ with $n\geq 3$, but it is logarithmic in $d=d_{n=2}=4$.
If the analytic continuation is performed anyway, there is thus an additional ``unbalanced'' singularity
which is seen as an additional $1/(n-2)$ pole in the beta functions.

\subsection{$\phi^{2n}$ universality class}\label{subsect:phi2n}

The leading quadratic contribution to the path integral \eqref{eq:leading-path-integral-contribution} is not
a one loop contribution for all $n\geq 3$ models, but rather it involves $(r-1)$-loops,
which is a marked distinction from the more familiar analyses of
$\phi^4$ and Yang-Mills theory below the upper critical dimension $d=4$.
To highlight this fact let us consider the first element of this family, which is $\phi^6$ for $n=3$
and which has been already renormalized in curved space in \cite{Huish:1994fb}:
the leading contributions to the renormalization of the couplings come from two loop diagrams
and in general contributions come from every other loop order \cite{Huish:1995cn}.

\begin{figure}[htb]
\includegraphics[width=0.4\textwidth]{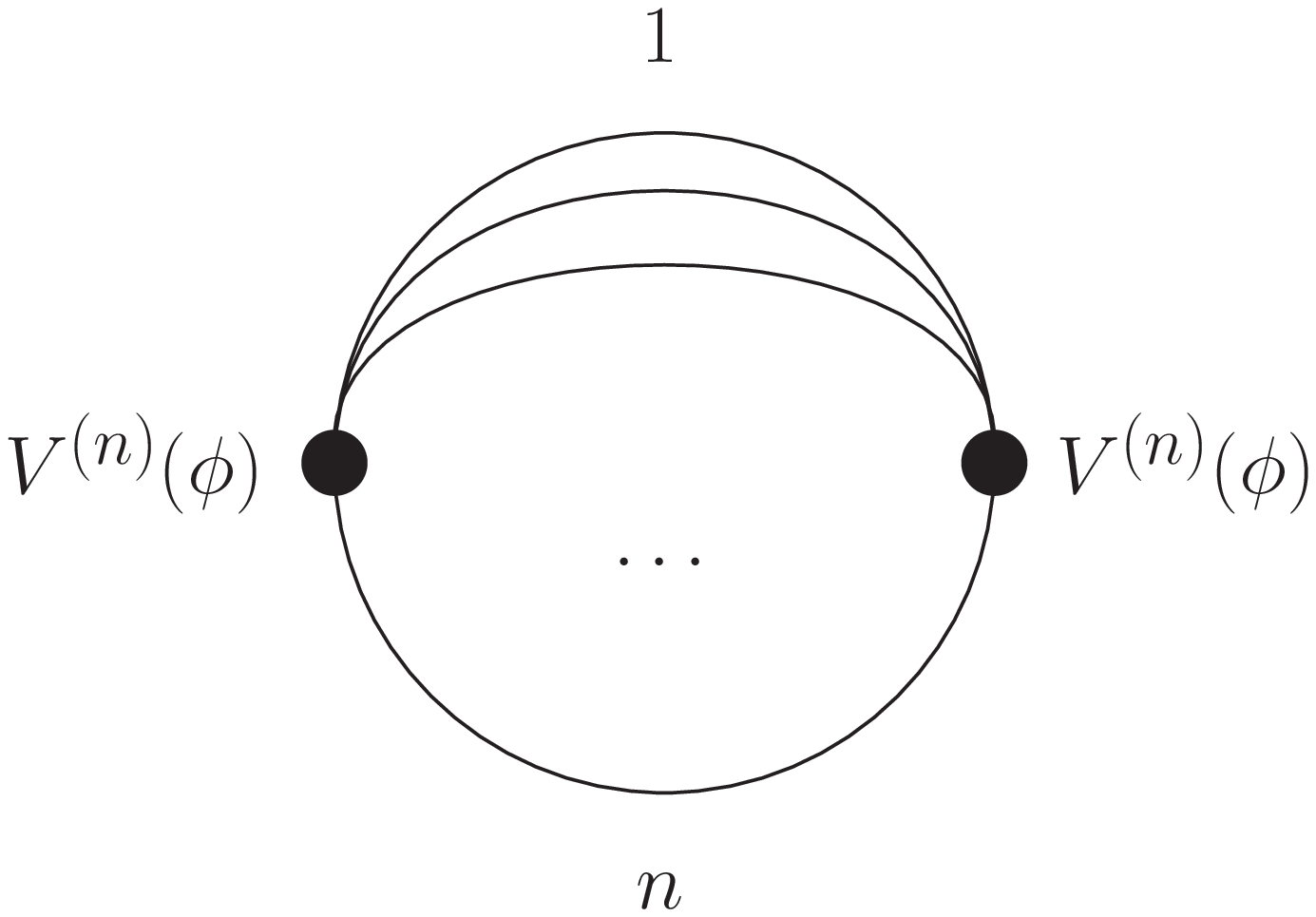}\\
\includegraphics[width=0.4\textwidth]{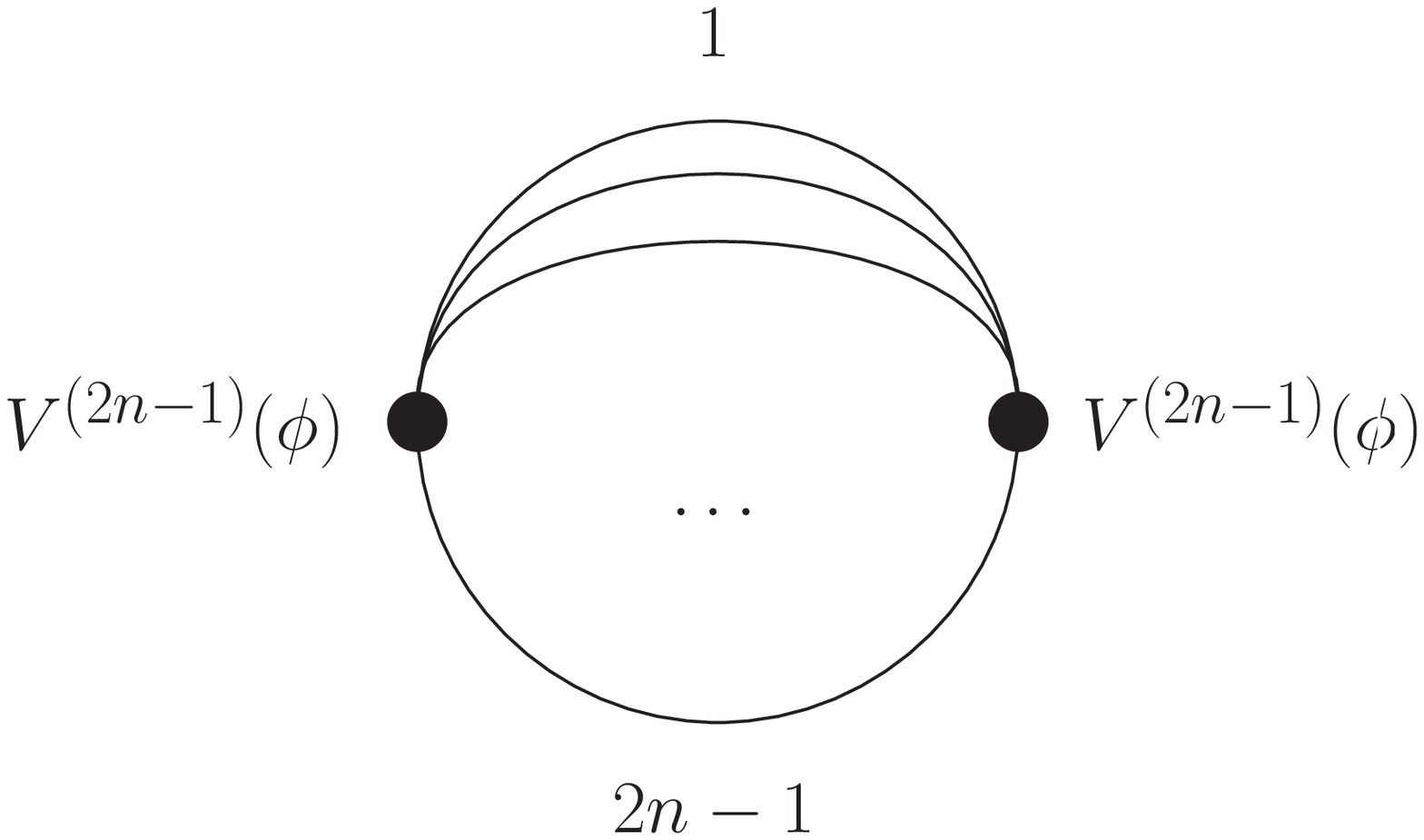}\\
\includegraphics[width=0.4\textwidth]{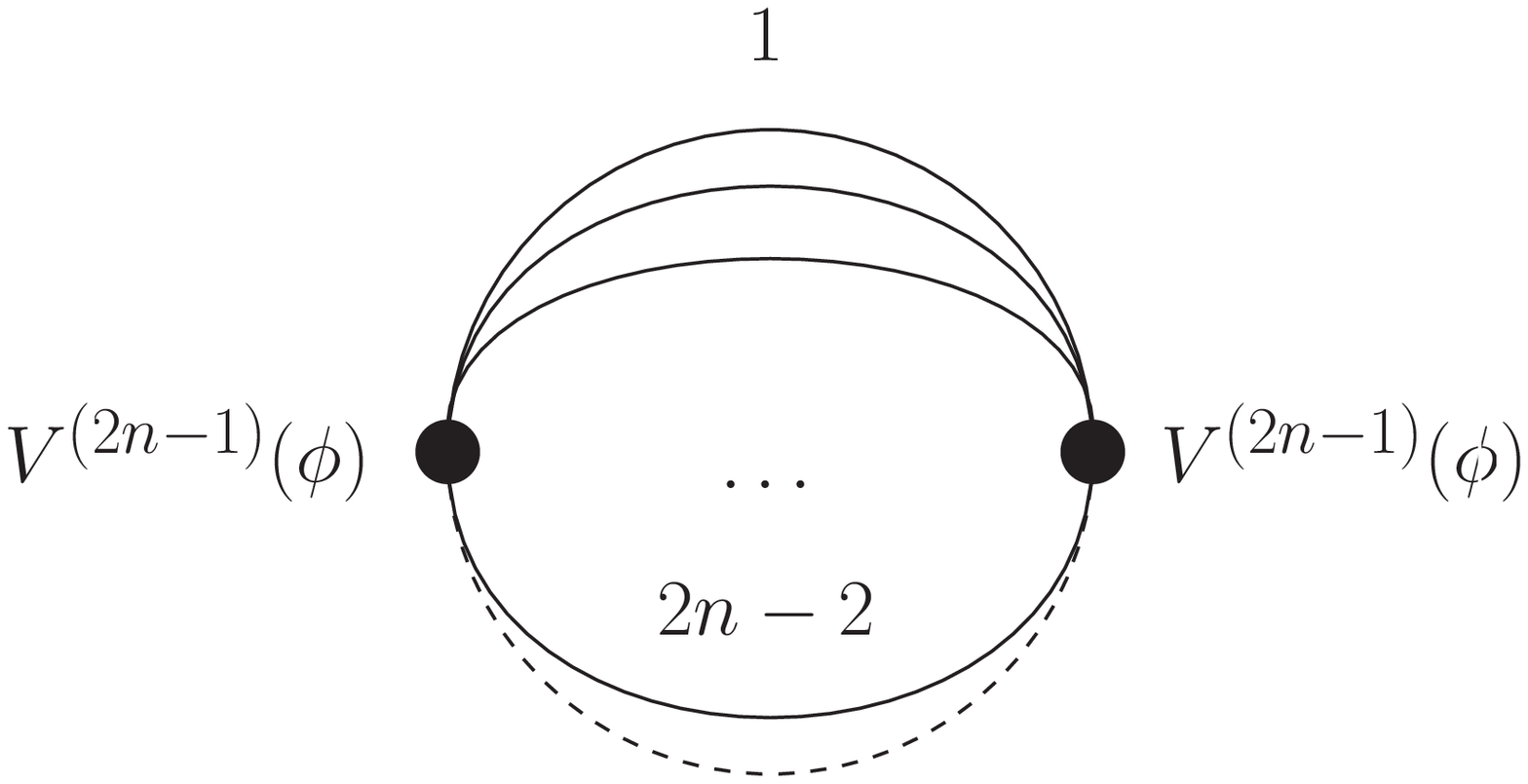}
\caption{Diagrammatic representation of \eqref{eq:counterterms} in order of appearance.
 The first and second diagrams are made of $n$ and $2n-1$ lines of the leading contribution of the Green function $G_0(x,x')$.
 Trivially their symmetry factors are $n!$ and $(2n-1)!$ respectively. The third diagram is again made of $2n-1$ lines,
 but one corresponds to the subleading $G_1(x,x')$ line which is depicted as dashed. Its symmetry factor is
 $(2n-2)!$ because there are $(2n-1)$ ways to choose the last line.
 }
\label{fig:diagrams}
\end{figure}

In general, not all loop contributions to \eqref{eq:leading-path-integral-contribution} lead to
$1/\epsilon$ poles for all values of $n$.
Using the methods described in Appendix \ref{sect:feynman} and dimensional analysis,
it is possible to infer that $1/\epsilon$ poles arise
for the cases $r=n$ and $r=2n-1$, corresponding to $(n-1)$- and $(2n-2)$-loop diagram respectively likewise flat space \cite{ODwyer:2007brp,Codello:2017hhh}.
In the case $r=n$, the contribution arises solely from $r$ lines of the leading $G_0(x,x')$ term of the Green function.
In the second case the diagram can be either composed by $2n-1$ lines of $G_0(x,x')$, or by $2n-2$ lines of $G_0(x,x')$ and one of $G_1(x,x')$.
In practice, this makes for three multiloop diagrams that must be evaluated by the methods described in Appendix \ref{sect:feynman}.
The diagrams are depicted in Fig.~\ref{fig:diagrams}.
We have that in $d=d_n-\epsilon$ the three diagrams evaluate to
\begin{widetext}
\begin{equation} \label{eq:counterterms}
\begin{split}
 & \frac{1}{2 n!} \int
 V^{(n)}(\phi)~ G_0^n ~V^{(n)}(\phi') 
 \sim
  c_n^{n-1} \mu^{(1-n)\epsilon} \frac{1}{4n!~\epsilon} V^{(n)}(\phi)^2 \\[0.5cm]
 & \frac{1}{2 (2n-1)!} \int
 V^{(2n-1)}(\phi)~ G_0^{2n-1} ~V^{(2n-1)}(\phi')
  %
 \sim -c_n^{2n-2}\mu^{2(1-n)\epsilon}\frac{(n-1)}{16(2n)!~\epsilon}\int \Bigl\{
 V^{(2n)}(\phi)^2 (\partial\phi)^{2} - \frac{2n-3}{6} V^{(2n-1)}(\phi)^2R
 \Bigr\} \\[0.5cm]
 & \frac{1}{2 (2n-2)!} \int
 V^{(2n-1)}(\phi)~ G_0^{2n-2}G_1 a_1 ~V^{(2n-1)}(\phi')
  %
  \sim c_n^{2n-2}\mu^{2(1-n)\epsilon} \frac{n(n-1)(2n-1)}{16(n-2) ~ (2n)!\epsilon} \int \Bigl\{ F''(\phi) -\frac{1}{6}\Bigr\}V^{(2n-1)}(\phi)^2R
\end{split}
\end{equation}
\end{widetext}
in which we suppress several coordinate indices on the left hand side for brevity.
We integrated by parts one derivative to cast the kinetic-like term of the second diagram in a suitable form, and defined the constant
\begin{equation}
\begin{split}
 c_n= \frac{1}{4\pi} ~\frac{1}{\pi^{\frac{1}{n-1}}} ~ \Gamma\left(\frac{1}{n-1}\right)\,.
\end{split}
\end{equation}
The results of \eqref{eq:counterterms} are essentially the counterterms which must be inserted
to remove the divergences of all the relevant operators of the $\phi^{2n}$ model in curved space for $n\geq 2$. The pole at $n=2$
of the last counterterm is a clear indication of why we left the $\phi^4$ models out of this general discussion.

Since we are just considering a leading renormalization,
the computation of the renormalization group flow is straightforward
because it can be obtained by simply acting on the counterterms with the logarithmic derivative
with respect to the reference scale $\mu\frac{\partial}{\partial\mu}$.
Naturally, we display the RG in the guise of functional equations.
We also include a field dependent wavefunction $Z(\phi)$ as renormalization of the kinetic term.
The wavefunction is generated by the flow and, while it includes irrelevant contributions for the most part,
the use of a boundary condition for $Z(0)$ allows for the determination of the anomalous dimension
of the renormalized field.
At the upper critical dimension we find
\begin{equation}\label{eq:phi-2n-functional-betas-dimful}
\begin{split}
 \beta_V =& \frac{c_n^{n-1}(n-1)}{4 ~ n!} V^{(n)}(\phi)^2\,, \\
 \beta_Z =& -\frac{c_n^{2n-2}(n-1)^2}{4 ~ (2n)!} V^{(2n)}(\phi)^2\,, \\
 \beta_F =& -\frac{c_n^{2n-2}(n-1)^2}{8(n-2)~ (2n)!}\Bigl\{
 (n-1) \\& - n (2n-1)F''(\phi)
 \Bigr\}V^{(2n-1)}(\phi)^2 \,.
\end{split}
\end{equation}

In a rather standard fashion we switch to the dimensionless renormalized canonically-normalized field
\begin{equation}
\begin{split}
 \varphi &= Z_0^{\frac{1}{2}} \mu^{-\frac{d-2}{2}} \phi\,,
\end{split}
\end{equation}
which includes a rescaling by the wavefunction renormalization constant $Z_0 =Z(0)$ which is generated by $\beta_Z$.
The field $\varphi$ is the natural argument for the dimensionless renormalized functions
$v(\varphi) = \mu^{-d} V(\phi)$, $z(\varphi)=Z_0^{-1} Z(\phi)$ and $f(\varphi)=\mu^{2-d} F(\phi)$.
Their renormalization group flow is
\begin{eqnarray}\label{eq:phi-2n-functional-betas}
 \beta_v &&= -d v +\frac{d-2+\eta}{2} \varphi v'+ \frac{c_n^{n-1}(n-1)}{4 ~ n!} (v^{(n)})^2\,, \nonumber \\
 \beta_z &&= \eta z + \frac{d-2+\eta}{2} \varphi z'-\frac{c_n^{2n-2}(n-1)^2}{4 ~ (2n)!} (v^{(2n)})^2\,, \\
 \beta_f &&= (2-d)f +\frac{d-2+\eta}{2} \varphi f' \nonumber \\ && -\frac{c_n^{2n-2}(n-1)^2}{8(n-2)~ (2n)!}\Bigl\{
 (n-1) - n (2n-1)f''
 \Bigr\}(v^{(2n-1)})^2 \,. \nonumber
\end{eqnarray}
By construction we have $z(0)=1$, so the limit $\varphi\to0$ can be used to determine the anomalous dimension
$\eta \equiv-\partial\log Z_0/\partial\log\mu$ directly from $\left.\beta_z\right|_{\varphi=0}=0$.

\subsection{$\phi^4$ universality class}\label{subsect:phi4}

The four dimensional case is special for three main reasons.
Firstly, diagrams and counterterms leading to the renormalization are not directly obtained as the analytic continuations
to $n=2$ of the results of Sect.~\ref{subsect:phi2n}.
Secondly, the subleading correction to the Green function in four dimensions is logarithmic.
This means that in the $\epsilon$-expansion an additional divergence must be subtracted from the propagator
as we show in \eqref{eq:G1-definition-4d}. The difference in the behavior of the subleading part of the propagator
is the reason why a $\frac{1}{n-2}$ pole appears in the third diagram of \eqref{eq:counterterms}.
Thirdly, a simple dimensional analysis reveals that operators quadratic in the curvatures
have the same canonical dimension of the operators $\phi^4$ and $\phi^2 R$,
and hence must be renormalized together for consistency.

Here we try to follow the notation of \cite{Jack:1983sk} for the most part with some minor modification.
Let us first generalize the action \eqref{eq:bare-action} to accommodate the higher curvatures
\begin{equation}\label{eq:bare-action-4d}
\begin{split}
 S[\phi] =& \int {\rm d}^dx \sqrt{g}\Bigl\{
 \frac{1}{2}g^{\mu\nu}\partial_\mu\phi\partial_\nu\phi +V(\phi) +F(\phi) R\\& - a \, {\cal F}  - b \, {\cal G} -c \, R^2 -e \, \nabla^2R
 \Bigr\}\,,
\end{split}
\end{equation}
with the following invariants
\begin{equation}
\begin{split}
 {\cal F} &= \frac{2}{(d-2)(d-1)}R^2-\frac{4}{d-2}R_{\mu\nu}R^{\mu\nu}+R_{\mu\nu\rho\theta}R^{\mu\nu\rho\theta}\,, \\
 {\cal G} &= R^2-4R_{\mu\nu}R^{\mu\nu}+R_{\mu\nu\rho\theta}R^{\mu\nu\rho\theta}\,.
\end{split}
\end{equation}
These invariants are chosen so that in four dimensions ${\cal G}$ integrates to a topological invariant
and ${\cal F}$, which is the square of the Weyl tensor, transforms covariantly under scale transformations.

It is convenient to define one general function and its modification as follows
\begin{equation}
\begin{split}
 U(\phi,R) &= V(\phi) +F(\phi) R - a {\cal F}  - b {\cal G}-c R^2-e \nabla^2R\,, \\  \hat{U}(\phi,R) &= U(\phi,R) -\frac{1}{12} R \phi^2\,.
\end{split}
\end{equation}
At one loop, which is the leading order, the counterterm to $U(\phi,R)$ can be obtained by a simple
application of the heat kernel. One finds that the leading contribution to the renormalization of $U(\phi,R)$
comes from the $a_2(x,x)$ coefficient given in \eqref{eq:hk-coincidence-limits}
\begin{equation}
\begin{split}
 -\frac{\mu^{-\epsilon}}{(4\pi)^2~\epsilon} \int \Bigl\{
  \frac{1}{2} \partial^2_\phi\hat{U}(\phi,R)^2 + \frac{1}{120} {\cal F} -\frac{1}{360} {\cal G}
 \Bigr\}\,,
\end{split}
\end{equation}
while the wavefunction renormalization is a two loop effect completely analog to the limit $n=2$ of Sect.~\ref{subsect:phi2n}.
The computation of the leading beta function is straightforward
\begin{equation}
\begin{split}
 \beta_U &= \frac{1}{(4\pi)^2} \Bigl\{ \frac{1}{2} \partial^2_\phi\hat{U}(\phi,R)^2 + \frac{1}{120} {\cal F} -\frac{1}{360} {\cal G} \Bigr\} \,, \\
 \beta_Z &= -\frac{1}{6(4\pi)^4} V^{(4)}(\phi)^2\,.
\end{split}
\end{equation}
Returning to the original functions of \eqref{eq:bare-action-4d} we find the functional beta functions
\begin{align}\label{eq:phi-4-functional-betas}
 &\beta_V = \frac{1}{2(4\pi)^2} V''(\phi)^2\,, \qquad 
 \beta_Z = -\frac{1}{6(4\pi)^4} V^{(4)}(\phi)^2\,,\nonumber\\
 &\beta_F = -\frac{1}{(4\pi)^2}\Bigl\{
 \frac{1}{6}-F''(\phi)
 \Bigr\}V''(\phi) \,.&&
\end{align}
as well as the beta functions for the higher derivative couplings
\begin{align}\label{eq:phi-4-hd-betas}
 &\beta_a = -\frac{1}{120(4\pi)^2}\,,   && \beta_c = \frac{1}{2(4\pi)^2}\Bigl\{\frac{1}{6}-F''(\phi)\Bigr\}^2 \,,\nonumber\\
 &\beta_b = \frac{1}{360(4\pi)^2}\,,  &&\beta_e = -\frac{1}{6(4\pi)^2}\Bigl\{\frac{1}{5}-F''(\phi)\Bigr\}\,.
\end{align}
Since $F(\phi)$ is at most quadratic we have that $F''(\phi)=F''(0)$ and the couplings $c$ and $e$ can be treated as numbers,
even though the right hand side suggests otherwise.

In order to make this section on $\phi^4$ more self consistent, we briefly discuss some critical property of the above system.
This discussion anticipates some points that are made later in the development of Sect.~\ref{sect:criticality}.
One can see that at the leading order the critical value for the nonminimal coupling $\xi=F''(0)$ is $\xi=\frac{1}{6}$
as one would naively expect from continuing the general conformal value \eqref{eq:xi-conformal} to $d=4$,
thus proving the educated guess of the introduction for the special case $n=2$.

In general it is not guaranteed that the critical value of $\xi$ remains a fixed point beyond the leading order
unless a further renormalization condition is exploited \cite{Brown:1980qq}.
For our purpose, it would be interesting to know if the general conjecture of the introduction is true,
that is, we would like to know under which circumstances at two loops and for $d=4-\epsilon$ the coupling takes the value
\begin{equation}\label{eq:fixed-point-xi-4d}
\begin{split}
 \xi=\frac{d-2}{4(d-1)} &= \frac{1}{6} -\frac{1}{36}\epsilon +\dots\,.
\end{split}
\end{equation}
One can prove, using naively the dimensionally regulated scheme at the next-to-leading order (NLO) and a straightforward subtraction,
that the above value is \emph{not} a fixed point to order $\epsilon$.
However, the freedom highlighted in \cite{Jack:1983sk} of redefining the potential $U(\phi,R)$ by a copy of the one loop counterterms
can be exploited to ensure that \eqref{eq:fixed-point-xi-4d} is the fixed point at NLO.
The redefinition is a change of the renormalization conditions which thus defines and links the metric and the field.
We refer to \cite{Jack:1983sk} for a more complete and detailed explanation of the results reported in this section.

\subsection{$\phi^\infty$ universality class: the Sine-Gordon model}\label{subsect:sg}

The upper critical dimension $d=2$ emerges as the limit $d_n\to 2$ of $n\to\infty$.
The renormalization of the path integral for the two dimensional case is very simple,
even though it represents a special case likewise the $\phi^4$ one.
It is convenient to borrow the notation from the previous section and use the full potential $U(\phi,R)$.
The computation of the leading counterterms and beta functions necessitates only the use of the standard
heat kernel expansion of an operator of Laplace-type, and specifically of the coefficient $a_1(x,x)$ given in \eqref{eq:hk-coincidence-limits}.
We find the leading counterterm at one loop
\begin{equation}
\begin{split}
 \frac{\mu^{-\epsilon}}{4\pi~ \epsilon} \int
  \partial_\phi^2 \hat{U}(\phi,R)
 \,,
\end{split}
\end{equation}
and deduce the very simple RG beta functional 
\begin{equation}\label{eq:beta-U-2d}
\begin{split}
 \beta_U=-\frac{1}{4\pi}\partial_\phi^2 \hat{U}(\phi,R)\,.
\end{split}
\end{equation}
Notice that there is no anomalous dimension renormalization coming from our leading order computation.

In two dimensions the scale invariant solutions of this beta function become periodic.
It has been argued that the critical solution of this RG flow in flat space is periodic and
corresponds to the Sine-Gordon universality class \cite{Codello:2017hhh}.
Here we are observing a generalization to curved spacetime for zero anomalous dimension as in \cite{Merzlikin:2017zan}.
Let us first introduce the dimensionless potential $u(\varphi,R)= \mu^{-2}U(\varphi,\mu^2 R)$.
Using the boundary conditions $u(\varphi,R)=u(-\varphi,R)$ and $\partial_\phi^2 U(\phi,R)|_{\phi=0}=m^2$,
at the fixed point in $d=2$ we find
\begin{equation}
\begin{split}
 u(\varphi,R) &= -\frac{m^2}{8\pi}\cos\left(\sqrt{8\pi}\varphi\right)+ \frac{R}{48\pi}\,.
\end{split}
\end{equation}
Notice that we have imposed the boundary conditions as a function of the scalar curvature, therefore
an implicit dependence on $R$ might in principle be hidden in the mass $m^2=m^2(R)$. In this way we have ensured
that the result agrees both with the assumption that this solution generalizes the Sine-Gordon universality to curved space,
and with the expectation that the nonminimal coupling $\xi$ should be zero at the critical point.

\subsection{$2d$ gravity at large-$c$}\label{subsect:2dgravity}

As a brief intermezzo we believe that it is interesting to show the relevance of the results of Sect.~\ref{subsect:sg}
in reproducing some well-known result of two dimensional quantum gravity coupled to conformal matter.
Let us recall that in exactly two dimensions the path integral of gravity can be determined by integrating the conformal anomaly \cite{Polyakov:1981rd},
which leads to a renormalization procedure linked to a nonlocal action known as the Polyakov action \cite{Distler:1988jt}.
This action is especially relevant because the spacetime integral of the Einstein term is a topological invariant in two dimension,
and hence it cannot govern the dynamic of the model.

However for general $d$ (and specifically for $d=2-\epsilon$) the Einstein term is not a topological invariant,
and therefore it has been argued by Kawai and Ninomiya that it should be possible to reproduce the results based on the Polyakov action
by just renormalizing the Einstein action in $d=2-\epsilon$ and then taking the limit $\epsilon\to 0$ \cite{Kawai:1989yh}.
The validity of this argument was shown through the course of several papers, which ultimately lead to the two loop renormalization
of the Einstein action in $d=2-\epsilon$. For more details we refer to \cite{Aida:1996zn} and references therein;
notice however that in the literature of $2d$ gravity it is often chosen $d=2+\epsilon$,
therefore the replacement $\epsilon\to-\epsilon$ is necessary when comparing results.

The renormalization of dimensionally regulated two dimensional gravity is slightly unconventional because it has to deal
with the conformal factor of the metric, otherwise one finds discontinuities when analytically continuing to $\epsilon\to0$ \cite{Jack:1990ey}.
In order to describe it, let us first introduce the Einstein action interacting with $c$ distinct conformally coupled fields $\phi_i$ in $d$ dimensions
\begin{equation}\label{eq:2d-gravity-action-1}
\begin{split}
 S[g,\phi] =& \int {\rm d}^dx \sqrt{g}\Bigl\{ -\frac{1}{G}R \\&+ \frac{1}{2}\sum_i\Bigl(\partial_\mu\phi^i\partial^\mu \phi^i + \xi_c \phi^i\phi^i R\Bigr) \Bigr\}\,.
\end{split}
\end{equation}
We require that the coupling $\xi_c$ is determined by the conformal value \eqref{eq:xi-conformal} and assume that this condition can be preserved through renormalization
(see the discussion of sections \ref{sect:introduction} and \ref{sect:criticality} for more details on this point).
The number $c$ is often referred to as ``central charge'' and it counts the effective number of matter degrees of freedom.

In two dimensions all possible metrics are related by a Weyl transformation, and therefore only their conformal mode is allowed to fluctuate.
Close to two dimensions, instead, it is customary to parametrize the metric $g_{\mu\nu} \to (\epsilon/8)^{2/\epsilon}\psi^{4/\epsilon} g_{\mu\nu}$ into a conformal mode $\psi$
and a metric $g_{\mu\nu}$ which is not allowed to fluctuate in its trace part (by abuse of notation we denote the transformed metric with $g_{\mu\nu}$).\footnote{
We find that the best recent review of this formulation appeared in \cite{Gielen:2018pvk}, in which
it has been named unimodular Dirac gravity, or alternatively unimodular dilaton gravity.
The gauge group $Diff^*$ of the formulation comes from the breaking of a semidirect product of diffeomorphisms and Weyl transformations
which is itself isomorphic to the diffeomorphisms group $Diff \ltimes Weyl\to Diff^*\simeq  Diff$, but acts on $\psi$ and $g_{\mu\nu}$ in a nonstandard way.
}
Using this normalization the mode $\psi$ of the metric enjoys a Weyl invariant action, which is in form analog to any of
those of the fields $\phi_i$, if not for an overall negative sign which makes $\psi$ an unstable ``scalar'' degree of freedom.
The idea of \cite{Aida:1994zc} is to transform \eqref{eq:2d-gravity-action-1} into 
\begin{equation}
\begin{split}
 S[g,\psi,\phi] =& \int {\rm d}^dx \sqrt{g}\Bigl\{- \frac{1}{G} L(\psi,\phi_i)R \\& -\frac{1}{2}\partial_\mu\psi\partial^\mu\psi + \frac{1}{2}\sum_i\partial_\mu\phi^i\partial^\mu \phi^i  \Bigr\}\,.
\end{split}
\end{equation}
and renormalize it such that the function $L(\psi,\phi_i)$ respects the conformal coupling.
The new function is normalized by $L(0,0)=1$, which is a necessary condition to read off the value of the Newton constant $G$.

Assuming that the instability of $\psi$ can be cured by opportunely Wick rotating the theory, it is possible to neglect the effects of the dilaton
field $\psi$ as compared to those of the multiplet $\phi_i$;
moreover one can argue that for large values of $c$ the fluctuations induced by the fields $\phi_i$ dominate over those of $g_{\mu\nu}$ too.
In other words, for large-$c$ it should be necessary to integrate only the loops of $\phi_i$,
but this is exactly the multifield generalization of what we have done in section \ref{subsect:sg} upon the identification
\begin{equation}\label{eq:G-L-relation}
\begin{split}
 -\frac{\mu^{-\epsilon}}{G} L(\phi_i)R = U(\phi_i,R)\,
\end{split}
\end{equation}
for the dimensionless versions of $L$ and $G$. Notice that in the large-$c$ limit we are dropping any parametric dependence on the mode $\psi$
to highlight the connection with the previous section.

Now we use \eqref{eq:G-L-relation} inside \eqref{eq:beta-U-2d} to determine the renormalization group flow of the renormalized $G$ and $L(\phi_i)$.
In order to separate the two beta functions we have to impose that $L(0)=1$ along the flow.
Additionally we impose that all fields $\phi_i$ are coupled
in the same way so that it will be sufficient to denote each one of them by $\phi$.
We find
\begin{equation}
\begin{split}
 \beta_G &= -\epsilon G +\frac{c}{24\pi} G^2+\frac{c}{4\pi}G \,L''(0)\\
 \beta_L &= -\frac{c}{24\pi}G\Bigl\{1-L(\phi)\Bigr\} +\frac{c}{4\pi}\Bigl\{L(\phi) ~L''(0) - L''(\phi)\Bigr\}\,.
\end{split}
\end{equation}
The interaction with the fluctuating modes of $g_{\mu\nu}$ can change the anomalous dimension of the fields $\phi_i$ as $\eta \sim G$,
but this contribution is also generally subleading in the limit of large central charge.
We follow the strategy of \cite{Aida:1994zc} and parametrize $L=1+a\psi +b\psi^2 -\xi_c \phi^2$.
It is straightforward to see that the beta functions of $a$ and $b$ have Gaussian solutions,
thus setting all couplings except for $G$ at the respective fixed points we obtain
\begin{equation}
\begin{split}
 \beta_G &= -\epsilon G + A G^2
\end{split}
\end{equation}
with $A= -\frac{c}{24\pi}$. This result agrees with the large-$c$ limit of the exact leading result
in which the constant $A$ takes the value $A=\frac{25-c}{24\pi}$ \cite{Kawai:1989yh}.

Notice that the general Euclidean result hinges on our ability of solving
the problem of the instability of the conformal mode, which in \cite{Aida:1994zc} is ``Wick'' rotated $\psi\to{\rm i}\psi$.
While several solutions have been proposed there is no definite answer,
nor general consensus on how to approach the problem.
In fact, proposals to solve the problem without a Wick rotation of the dilaton mode
have received renewed attention recently \cite{Morris:2018mhd}.
This problem can be framed in the more general discussion of finding the universality class of quantum gravity
and exploring the corresponding conformal theory \cite{Nink:2015lmq}.
Here we would like to mention another less explored yet interesting possibility that was outlined in \cite{Codello:2011yf}:
the path integral of $2d$ gravity could be ``defined''
starting with the path integral of a fluid $2d$ membrane embedded in $D$ bulk dimensions (which is essentially a non-critical Nambu-Goto string)
and analytically continuing to $D\to 0$. In the membrane path integral the correct counting
of the degrees of freedom involves the propagation of modes of the extrinsic curvature,
which play a role analog to the gauge fixing ghosts.

\subsection{$\phi^{2n+1}$ universality class}\label{subsect:phi2n+1}

The results of Sect.~\ref{subsect:phi2n} can in part be generalized to
the tower of multicritical \emph{non-unitary} models $\phi^{2n+1}$ \cite{Codello:2017epp}.
The first model of such tower would be $\phi^3$ whose curved-space renormalization
has been studied in \cite{Toms:1982af}.
While it can be seen that the quadratic leading term in the renormalization of
the potential is replaced by a more involved cubic one,
it turns out that the leading renormalization of the wavefunction and the non-minimal coupling function
is contained in the same diagrams with the opportune change of the number of internal lines.

A simple rule of thumb to test the validity of the functional RG equations
involves the replacement $n\to n+\frac{1}{2}$ in the system \eqref{eq:phi-2n-functional-betas}:
the beta function $\beta_v$ ceases to make sense signaling that it should be replaced with a cubic term,
but both $\beta_z$ and $\beta_f$ are still meaningful and in fact they are the correct beta functions.
If the leading cubic flow of the potential, which is given in \cite{Codello:2017epp}, is included,
then it is trivial to generalize the system \eqref{eq:phi-2n-functional-betas} to the entire sequence of $\phi^{2n+1}$ models.

We checked explicitly that in the special case $\phi^3$ the resulting system,
which corresponds to the Lee-Yang universality class with a nonminimal coupling to the curvature,
coincides with the one given in the appendix of \cite{Merzlikin:2017zan}.
Notice that the cubic case should be treated with more care than we do, because the complete renormalization
requires counterterms for all the cubic invariants coming from the metric \cite{Jack:1985wf},
and therefore it is as special as the quartic case described in Sect.~\ref{subsect:phi4}.
The general next-to-leading renormalization (occurring at two loops) of $\phi^3$ in curved space
appeared for the first time in \cite{Jack:1985wf}.
The renormalization of the model is actually known with position-dependent couplings \cite{Grinstein:2015ina},
a result which is used to show that a natural six dimensional generalization of the Zamolodchikov's $c$-function
does \emph{not} always increase monotonically with the flow \cite{Grinstein:2014xba}.

\section{Criticality}\label{sect:criticality}

We now resume the analysis of the RG system \eqref{eq:phi-2n-functional-betas} of Sect.~\ref{subsect:phi2n} representing the general case
of the multicritical model $\phi^{2n}$ for $n\ge 3$. The $\phi^4$ model is an outlier,
so we anticipated a brief discussion of the critical properties of its nonminimal coupling to the curvature already in Sect.~\ref{subsect:phi4}.
We find convenient to rescale the potential
\begin{equation}
\begin{split}
 v(\varphi) \to \frac{4 ~ c_n^{1-n}}{n-1} v(\varphi) = \frac{(4\pi)^n}{n-1} \Gamma\left(\frac{1}{n-1}\right)^{1-n} v(\varphi)\,,
\end{split}
\end{equation}
while leaving all other functions intact.
The system \eqref{eq:phi-2n-functional-betas} simplifies to
\begin{equation}\label{eq:phi-2n-functional-betas-rescaled}
\begin{split}
 \beta_v =&
  -d v +\frac{d-2+\eta}{2} \varphi v'+ \frac{1}{n!} (v^{(n)})^2\,, \\
 \beta_f =& (2-d)f +\frac{d-2+\eta}{2} \varphi f' \\& -\frac{2n (2n-1)}{(n-2)~(2n)!}\Bigl\{
 \frac{n-1}{n (2n-1)} - f''
 \Bigr\}(v^{(2n-1)})^2 \,.
\end{split}
\end{equation}
Using the boundary condition $z(0)=1$ in the rescaled flow $\beta_z$, we also determine the anomalous dimension of the scalar field $\eta= 4 v^{(2n)}(0)^2/(2n)!$.

The critical couplings appear as the leading couplings of the potentials $v(\varphi)$ and $f(\varphi)$.
By construction, in the minimal subtraction scheme all other couplings are dimensionful, and therefore are zero at the critical point.
We therefore parametrize the potentials in terms of the two almost marginal interactions
\begin{equation}
\begin{split}
 v(\varphi) = \frac{\lambda}{(2n)!}\varphi^{2n}\,, \qquad \qquad f(\varphi) = \frac{\xi}{2}\varphi^{2}\,.
\end{split}
\end{equation}
Using the above parametrization in \eqref{eq:phi-2n-functional-betas-rescaled},
we find the following beta functions and anomalous dimension
\begin{equation}\label{eq:beta-system}
\begin{split}
 &\beta_\lambda = -(n-1)\epsilon\lambda +\eta n \lambda + \frac{(2n)!}{(n!)^2}\lambda^2\,,
 \qquad \eta = \frac{4}{(2n)!}\lambda^2\,, \\
 &\beta_\xi = \eta\xi - \frac{4(n-1)}{(n-2)(2n)!}\lambda^2 + \frac{4n(2n-1)}{(n-2)(2n)!} \xi\lambda^2\,.
\end{split}
\end{equation}
It is clear that $\eta$ contributes to the cubic order in $\lambda$ of $\beta_\lambda$,
which has no effect to the determination of the order $\epsilon$ of the fixed point.
However $\eta$ has an important effect in $\beta_\xi$ because its contribution scales
with the same power of $\lambda$ as the other terms. Substituting $\eta$ we find
\begin{equation}\label{eq:betas-lambda-xi-final}
\begin{split}
 &\beta_\lambda = -(n-1)\epsilon\lambda + \frac{(2n)!}{(n!)^2}\lambda^2\,,
 \\
 &\beta_\xi = \frac{8(n^2-1)}{(n-2)~(2n)!}\left(\xi-\frac{1}{2(n+1)}\right)\lambda^2\,.
\end{split}
\end{equation}

The system has two different fixed points. On the one hand we have the Gaussian fixed point at
$\lambda=0$ which sets both beta functions to zero. In this case the natural fixed point for $\xi$ is the subleading
root of $\beta_\xi$. On the other hand we have the non Gaussian fixed point
\begin{equation}\label{eq:lambda-fp}
\begin{split}
 &\lambda^* = \frac{(n-1)~(n!)^2}{(2n)!}\epsilon\,,\qquad \qquad \xi^*= \frac{1}{2(n+1)}\,.
\end{split}
\end{equation}
For both fixed points the coupling $\xi$ takes the critical value that is expected \emph{at the upper critical dimension}
\begin{equation}\label{eq:xi-fp}
\begin{split}
\xi^* = \xi_n \equiv \frac{d_n-2}{4(d_n-1)}= \frac{1}{2(n+1)}\,,
\end{split}
\end{equation}
which evidently proves the educated guess given in the introduction.
The above analysis can be extended easily to the multicritical nonunitary models $\phi^{2n+1}$ following the guidelines
explained in Sect.~\ref{subsect:phi2n+1}.
Interestingly, the only outlier of our analysis is the case for $n=2$, for which we have to use the set of beta functions
coming from \eqref{eq:phi-4-functional-betas} as discussed in Sect.~\ref{subsect:phi4}.
However, it is straightforward to find that in this case $\xi=\frac{1}{6}$
which happens to coincide with the continuation of \eqref{eq:xi-fp} to $n=2$.
It is an easy check to see that the limit $n\to \frac{3}{2}$ in \eqref{eq:xi-fp} gives $\xi^*=\frac{1}{5}$ as shown in \cite{Toms:1982af}.

The next step would be to test if the next-to-leading order correction to the non-Gaussian fixed point of $\xi$
matches the $\epsilon$-expansion of conformal value for the coupling $\xi$ evaluated in $d=d_n-\epsilon$ instead of $d=d_n$
in agreement with the general conjecture of the introduction.
This would imply
\begin{equation}\label{eq:xi-fp-conjecture}
\begin{split}
 \xi \overset{?}{=} \frac{d-2}{4(d-1)}=\frac{1}{2(n+1)} - \frac{(n-1)^2}{4(n+1)^2}~ \epsilon + \dots \,,
\end{split}
\end{equation}
which comes from the expansion of \eqref{eq:xi-conformal} to orders of $\epsilon$ using $d=d_n-\epsilon$.
Following the discussion of \cite{Brown:1980qq,Jack:1983sk}, which we reproduced briefly in Sect.~\ref{subsect:phi4},
we argue that ensuring \eqref{eq:xi-fp-conjecture} probably requires a special choice in
the renormalization conditions leading the the RG flow. In practice, the
next-to-leading contributions to the RG flow can be changed by the inclusion of terms which match
the counterterms \eqref{eq:counterterms} and which can be used to change the renormalization conditions
leading to the fixed point value for $\xi$. We hope to come back to this topic.
Let us include here also a short remark on the steps that have lead to \eqref{eq:xi-fp}.
While our educated guess of the introduction stated that we expected that $\xi$ takes the conformal value at criticality,
the validity of the guess is not at all obvious from the initial form of the counterterms \eqref{eq:counterterms}.
In particular, there is a very delicate balance among the terms appearing in the renormalization \eqref{eq:phi-2n-functional-betas}
and the anomalous dimension which produces the form of $\beta_\xi$ in \eqref{eq:betas-lambda-xi-final}
and which makes evident that the conformal value \eqref{eq:xi-fp} is actually the critical point.

We conclude this section by discussing the implications that the system of beta functions \eqref{eq:betas-lambda-xi-final} has on the infrared physics.
For obvious reasons, we are mostly interested in studying the renormalization group flow in a physical dimension.
The first natural dimension (smaller than $d_n$) in which almost all models for $n\geq 3$ are nontrivial is $d=2$,
we therefore continue $\epsilon$ to the value $\epsilon = \frac{2}{n-1}$
to continue the $\phi^{2n}$ models to the physical dimension $d=2$.
Correspondingly, the fixed point value of the coupling $\lambda$ becomes $\lambda^*= \frac{2}{A_n}$
in which we define $A_n=(2n)!/(n!)^2$ which is simply the coefficient of the $\lambda^2$ term in $\beta_\lambda$.
The flow can be integrated as follows
\begin{equation}\label{eq:integrated-flow}
\begin{split}
 &\lambda(\mu) = \frac{\lambda_0}{\frac{\lambda_0}{\lambda^*}+\left(\frac{\mu}{\mu_0}\right)^2\left(1-\frac{\lambda_0}{\lambda^*}\right)} \,,
 \\&
 \xi(\mu) = \xi_n + (\xi_0-\xi_n) ~e^{-B_n\int_\mu^{\mu_0} d\rho\,\frac{\lambda^2(\rho)}{\rho}}\,,
\end{split}
\end{equation}
in which we introduce $B_n = 8(n^2-1)/((n-2)~(2n)!)$ that is the coefficient of $\beta_\xi$.
The flow satisfies the ultraviolet boundary conditions $\lambda_0=\lambda(\mu_0)$ and $\xi_0=\xi(\mu_0)$,
which can be checked by setting $\mu=\mu_0$ in \eqref{eq:integrated-flow}.

More interestingly, we can use \eqref{eq:integrated-flow} to explore the infrared limit $\mu=0$.
One can see trivially that the second term in the denominator of $\lambda(\mu)$ drops for $\mu=0$ and therefore we have $\lambda(0)=\lambda^*$.
Slightly less trivial is to show that for $\mu\to0$ the integral appearing in the exponential of $\xi(\mu)$ diverges logarithmically
implying that the second term drops; we thus have $\xi(0) = \xi_n$.
These results are in line with the expectation that the nontrivial fixed point \eqref{eq:lambda-fp}--\eqref{eq:xi-fp} is of infrared nature
in that it controls the large scale behavior of the model near criticality.

\section{Conclusions}\label{sect:conclusions}

We have considered the leading order renormalization of the multicritical scalar models with $\phi^{2n}$
interaction in curved space. Our analysis shows that for almost all values of $n$ one has to consider counterterms
for the self interaction as well as for the nonminimal interaction of the form $\phi^2 R$,
while some additional counterterms based on curvature invariants are needed in the special case $n=2$.
The counterterms have been obtained from a computation of the $\frac{1}{\epsilon}$ poles of dimensionally
regulated covariant Feynman diagrams of $(n-1)$-loops for the self interaction, and $(2n-1)$-loops for the self energy and the nonminimal interaction.

Our result generalizes the renormalization of the $\phi^4$ model in curved space, which we have considered as a special case,
but it also shows that the general case functions rather differently. Specifically, the structure
of the counterterms for the nonminimal coupling displays a discontinuity for $n=2$, which corresponds to $\phi^4$.
We have deduced a set of functional beta functions which describes the scale dependence of a
self-interaction potential and a generalized nonminimal interaction with the scalar curvature.

We have used the perturbative renormalization group flow to determine standard perturbative beta functions
for the two canonically marginal couplings: $\lambda$ of the self-interaction $\phi^{2n}$ and $\xi$ of the nonminimal interaction $\phi^2 R$.
The RG system clearly shows that at the leading order the scale invariant fixed point of the nonminimal coupling $\xi$
coincides with its dimension-dependent conformal value $\xi_c$ evaluated at the upper critical dimension.
This result is in agreement with an educated guess enunciated in the introduction.
Importantly, the leading critical value for the coupling $\xi$ is an ultraviolet attractive feature of the renormalization group flow.

We have also discussed a more general conjecture for which at the next-to-leading order
the $\epsilon$-expansion of the fixed point value of $\xi$ matches the expansion of the conformal value $\xi_c$
below the upper critical dimension. Based on similar and already available results for the $\phi^4$ \cite{Brown:1980qq}
and $\phi^3$ models \cite{Jack:1985wf,Grinstein:2014xba,Grinstein:2015ina},
we argue that one has to either follow a modified version of the prescription of Brown and Collins \cite{Brown:1980qq},
or alternatively to subtract normally while exploiting the freedom of redefining the renormalization
group flow at the next-to-leading order using the counterterms of the leading order \cite{ODwyer:2007brp}.
In other words, one might want to \emph{find} the appropriate renormalization condition
which ensures the validity of the conjecture for the nonminimal coupling.
We believe that this condition plays an important role,
especially if it is necessary to describe the model in a conformal or Weyl-invariant way.

A clearer understanding of the status of the conjecture and the renormalization condition
can only be achieved by building on the results of this paper and studying the next-to-leading order contributions
to the renormalization group flow, which is thus an important future prospect for this computation.

\bigskip

\noindent \emph{Acknowledgements.}
We are grateful to H.~Gies, R.~Percacci, I.~L.~Shapiro, G.~P.~Vacca and A.~Wipf for insights and discussions
which stimulated the analysis of this paper.
RM and OZ acknowledge support from the DFG under Grants Gi~328/7-1 and Za~958/2-1 respectively.


\appendix

\section{Covariant representation of the Green function}\label{sect:heat-kernel}

This and the next appendix follow roughly the presentation of \cite{Jack:1983sk} but summarize and adapt it to the purpose of this paper.
We restrict our attention to simple scalar fields, but the inclusion of internal indices and a gauge connection is straightforward.
Let us consider an operator of Laplace-type
\begin{equation}\label{eq:operator-form}
\begin{split}
 {\cal O} = - g^{\mu\nu}\nabla_\mu\partial_\nu + E\,,
\end{split}
\end{equation}
in which we included a local endomorphism $E=E(x)$ acting multiplicatively on the scalar field's bundle.
Notice that the spacetime metric $g_{\mu\nu}$ appears both through the inverse $g^{\mu\nu}$
and inside the Christoffel's symbols $\Gamma$ of Levi-Civita connection $\nabla=\partial+\Gamma$.

In the background field approach to the Euclidean path integral the curved space
propagator of the scalar field corresponds to the Green function of the operator ${\cal O}$
for an opportune choice of the endomorphism $E$. The Green function is defined as
\begin{equation}
\begin{split}
 {\cal O}_x G(x,x') = \delta^{(d)}(x,x')\,,
\end{split}
\end{equation}
in which we introduced the biscalar $\delta$-function that generalizes the usual flat space Dirac delta.
The propagator that is used in the main text can be obtained by specifying the endomorphism as $E=F''(\phi) R$.

It is convenient to represent the Green function using the heat kernel method.
The heat kernel function is defined as the solution of the following differential equation
\begin{equation}
\begin{split}
 &\partial_s {\cal G}(s;x,x') + {\cal O}_x {\cal G}(s;x,x') = 0\,,
 \\&
 {\cal G}(0;x,x') = \delta^{(d)}(x,x')\,.
\end{split}
\end{equation}
If we solve the diffusion equation implicitly
\begin{equation}
\begin{split}
  {\cal G}(s;x,x') = \langle x' | ~ {\rm e}^{-s {\cal O}} ~ | x\, \rangle \,,
\end{split}
\end{equation}
then the relation of the heat kernel function with the Green function is straightforward
\begin{equation}\label{eq:green-sdw-relation}
\begin{split}
  G(x,x') = \int_0^{\infty} {\rm d}s \, {\cal G}(s;x,x') \,.
\end{split}
\end{equation}
For all intents and purposes the above relation should be taken as our operative definition of $G(x,x')$.

The heat kernel representation is useful because the solution admits an asymptotic expansion
for small values of the parameter $s$, known as the Seeley-de Witt expansion,
which captures the ultraviolet properties of the Green function.
The expansion is generally parametrized as
\begin{equation}\label{eq:sdw-expansion}
\begin{split}
 {\cal G}(s;x,x') &= \frac{\Delta(x,x')^{1/2}}{(4\pi s)^{d/2}}  ~ {\rm e}^{-\frac{\sigma(x,x')}{2s}} \sum_{k\geq 0} a_k(x,x') ~ s^k\,.
\end{split}
\end{equation}
We introduced several bitensors in the expansion. The most fundamental is $\sigma(x,x')$, sometimes known as Synge's or Synge-de Witt's world function,
which is half of the square of the geodesic distance between the points $x$ and $x'$ \cite{J.L.Synge:1960zz}.
The bitensor $\Delta(x,x')$ is known as the van Vleck determinant
and is related to the world function and the determinant of the metric as
$$\Delta(x,x')=(g(x) g(x'))^{-1/2} \det \left(-\partial_\mu\partial_{\nu'} \sigma \right)\,.$$
Together, the bitensors $\sigma(x,x')$ and $\Delta(x,x')$ ensure that the leading term
of the Seeley-de Witt parametrization covariantly generalizes the solution of the heat equation in flat space with ${\cal O}\sim -\partial^2$.
Finally, the bitensors $a_k(x,x')$ are the coefficients of the asymptotic expansion
and contain the geometrical information of the operator ${\cal O}$, which includes
curvatures, connections and interactions.

It is well-known that ultraviolet properties are (and must be) local in renormalizable theories.
For the case of the heat kernel and the Green function locality corresponds to $x\sim x'$
and it is captured by the so-called coincidence limit in which $x\to x'$.
Given any bitensor $B(x,x')$, its coincidence limit is defined
\begin{equation}
\begin{split}
 [B] &= \lim_{x'\to x} B(x,x')\,.
\end{split}
\end{equation}
Notice that covariant derivatives do not generally commute with the coincidence limit $\nabla[B]\neq [\nabla B]$,
but rather satisfy a modified relation \cite{J.L.Synge:1960zz,Christensen:1976vb}.

The coincidence limits of the bitensors $\sigma(x,x')$ and $\Delta(x,x')$ and their derivatives can be obtained
by repeated differentiation of the \emph{crucial} relations
\begin{equation}\label{eq:basic-relations-bitensors}
\begin{split}
 \sigma_\mu \sigma^\mu = 2 \sigma\,,\qquad
 \Delta^{1/2} \sigma_\mu{}^\mu +2 \sigma^\mu \nabla_\mu \Delta^{1/2} = d \Delta^{1/2}\,,
\end{split}
\end{equation}
for which we suppressed bitensor coordinates and
we used the notation in which subscripts of $\sigma(x,x')$ correspond to covariant derivatives.
Similarly, coincidence limits of the coefficients $a_k(x,x')$ can be obtained
by differentiating and inductively using
\begin{equation}
\begin{split}
 & k a_k+\sigma^\mu \nabla_\mu a_k +\Delta^{-1/2} {\cal O} (\Delta^{1/2} a_{k-1}) =0
\end{split}
\end{equation}
with the boundary condition $\sigma^\mu\nabla_\mu a_0=0$. In the relevant example of a simple scalar field
the first coefficient is trivial $a_0(x,x')=1$, because the Seeley-de Witt expansion solves the diffusion equation in flat space.
We give here the first two nontrivial coincidence limits for the expansion of the operator \eqref{eq:operator-form} which are used in the computations of the main text
\begin{equation}\label{eq:hk-coincidence-limits}
\begin{split}
  \left[a_1\right] &= \frac{R}{6}-E \,,  \\
  \left[a_2\right] &= \frac{1}{72}R^2-\frac{1}{6}RE+\frac{1}{2}E^2
  -\frac{1}{6}\nabla^2\left(E-\frac{1}{6}R\right) \\&
  +\frac{1}{180}\left(R_{\mu\nu\rho\sigma}R^{\mu\nu\rho\sigma}-R_{\mu\nu}R^{\mu\nu}\right) \,. 
\end{split}
\end{equation}

Using \eqref{eq:sdw-expansion} in \eqref{eq:green-sdw-relation}, we obtain an analog expansion
of the Green function
\begin{equation}\label{eq:green-sdw-expansion}
\begin{split}
  G(x,x') = \sum_{k\geq 0} G_k(x,x') a_k(x,x')\,.
\end{split}
\end{equation}
The leading $G_0(x,x')$ and the subleading $G_k(x,x')$ for $k\geq 1$ are bilocal contributions to the Green function and
are determined by a simple integration over the heat kernel parameter $s$
\begin{equation}\label{eq:Gk-definition}
\begin{split}
  G_k(x,x') = \frac{2^{d-2-2k}}{(4\pi)^{d/2}}
  \frac{\Delta^{1/2}}{(2\sigma)^{d/2-1-k}} \Gamma\left(\frac{d}{2}-1-k\right)
  \,.
\end{split}
\end{equation}
Depending on the theory and its dimensionality, the exponent $d/2-1-k$
inevitably becomes negative for a certain value of $k$ highlighting the fact that there is only a finite number of
Green's function contributions which are singular in the limit $x \sim x'$.

It is very important to point out that in \eqref{eq:Gk-definition} we have implicitly assumed that the Green contributions
do not scale logarithmically with $\sigma(x,x')$ at the critical dimension.
While they do not scale logarithmically for almost all the multicritical
models considered in this paper (that is, for all $\phi^{2n}$ with $3\leq n < \infty$).
They do, however, scale logarithmically if there are values of $k$ for which $d=2+k$, which corresponds to poles of the gamma function.
In this case, one sees the failure of capturing the logarithmic behavior explicitly 
through the $\epsilon\to 0$ limit of \eqref{eq:Gk-definition} which is not regular even outside $x \sim x'$.

For this reason, when $d=2+k$ one needs to subtract an $\epsilon$-pole to \eqref{eq:Gk-definition}
for the results to be valid \emph{at and close to} the dimension $d$.
For example, close to $d=2$ the leading propagator is already logarithmic and we subtract
\begin{equation}\label{eq:Gk-definition-2d}
\begin{split}
  G^{2-\epsilon}_{0}(x,x') =
  \frac{\Delta^{1/2}}{(4\pi)^{d/2}} \frac{\Gamma\left(d/2-1\right)}{(2\sigma)^{d/2-1}} +  \mu^{-\epsilon} ~ \frac{ \Delta^{1/2}}{2 \pi \epsilon}
  \,.
\end{split}
\end{equation}
As desired, the above expression is valid for $d$ close to two dimensions and is regular in the limit $\epsilon\to 0$ for $d=2-\epsilon$.
We report here the subleading part of the Green function for $d=4-\epsilon$, which is also needed in the paper
\begin{equation}\label{eq:G1-definition-4d}
\begin{split}
  G^{4-\epsilon}_{1}(x,x') =
  \frac{\Delta^{1/2}}{(4\pi)^{d/2}} \frac{\Gamma\left(d/2-2\right)}{(2\sigma)^{d/2-2}} +  \mu^{-\epsilon} ~ \frac{ \Delta^{1/2}}{8 \pi^2 \epsilon} 
  \,.
\end{split}
\end{equation}
General expressions for leading and subleading parts can be found in \cite{Osborn:1987au,Jack:1983sk}.
The generalization to $d=6-\epsilon$, which requires subtractions starting from $k=2$, can be found in \cite{Jack:1985wf,Grinstein:2015ina}.

The need for a correct subtraction of the $\epsilon\to 0$ limit can be seen in the general counterterms \eqref{eq:counterterms}.
In fact the $(n-2)$ pole of the third counterterm is a symptom of the fact that the general $n\geq 3$ results
cannot be straightforwardly continued because in $d=4$ because the first subleading propagator becomes logarithmic.
This is the main reason why the renormalization of the $\phi^4$ universality class is an outlier.

\section{$\epsilon$-poles in curved space}\label{sect:feynman}

Covariant Feynman diagrams are constructed as products of propagators and hence of Green functions.
Taking advantage of the Seeley-de Witt representation given in Appendix \ref{sect:heat-kernel},
we notice that diagrams are generally written as products of powers of $\sigma(x,x')$, $\Delta(x,x')$, heat kernel coefficients
and eventually other bilocal operators which could be introduced by the theory's vertices.
In the case of a simple scalar field with a canonical kinetic term and no derivative interactions
the leading Feynman diagrams only have local vertices and can be represented by products of ``bundles''
of propagators. In the applications of this paper, there is always just one bundle
of propagators attached to the same two spacetime points as seen in \eqref{eq:counterterms}.

Generally, we want to obtain the dimensionally regulated divergent parts of covariant structures
of the form
\begin{equation}\label{eq:arbitrary-diagram}
\begin{split}
  Q(x,x') \Delta(x,x')^a \frac{1}{\sigma(x,x')^b}
\end{split}
\end{equation}
in which $Q(x,x')$ is an arbitrary bilocal operator, coming from the Seeley-de Witt coefficients, or the vertices, or other parts of the diagram.
The constants $a$ and $b$ are arbitrary powers that depend by the details of the diagram itself (for example by the number of propagators and the value of the critical dimension).
In the following, we shall briefly describe an algorithm due to Jack and Osborn which was developed to treat this kind of structures in dimensional regularization \cite{Jack:1983sk}.

One starts with the basic relation
\begin{equation}\label{eq:basic-divergence}
\begin{split}
  \frac{1}{\sigma(x,x')^{\frac{d}{2}-c\epsilon}} \sim  \frac{(2\pi)^{\frac{d}{2}}}{c ~\epsilon ~ \Gamma(d/2)} ~ \mu^{-2c\epsilon} ~\delta^{(d)}(x,x')\,,
\end{split}
\end{equation}
in which we introduced the symbol $\sim$ to establish equivalence of the divergent parts of both sides of the equation and a reference scale $\mu$
to preserve the dimensionality of the right hand side.
It is easy to prove the above relation in flat space for which it is sufficient to perform a Fourier transform and use the fact that $\sigma(x,x')=|x-x'|^2/2$;
it is then sufficient to argue that divergences are local and there cannot be curvature corrections on the right hand side because of dimensional reasons.
More generally, this relation can be proven using Riemann normal coordinates in curved space \cite{Jack:1983sk}.

Notice that if both sides of \eqref{eq:basic-divergence} are multiplied by the same bilocal operator,
then the Dirac delta on the right hand side allows for the substitution of its coincidence limit
$Q(x,x') \delta^{(d)}(x,x') = [Q]\delta^{(d)}(x,x')$.
The core of the algorithm is thus to transform all possible inverse powers of the world function into those of the left-hand-side of the basic relation,
substitute them with the right-hand-side,
and then sort all bilocal operators at the numerator so that they enter in contact with the Delta function.

Higher inverse powers of the world function can be manipulated inverting
\begin{equation}\label{eq:powers-of-sigma-relation}
\begin{split}
  & (\nabla^2-Y)\frac{\Delta^{1/2}}{\sigma^b} = b(2(b+1)-d) \frac{\Delta^{1/2}}{\sigma^{b+1}}\,,
  \\&
  Y(x,x')\equiv \Delta^{-1/2}\nabla^2\Delta^{1/2} \,,
\end{split}
\end{equation}
which can be proven easily using \eqref{eq:basic-relations-bitensors}. For the purpose of this paper we just need
\begin{equation}\label{eq:basic-divergence+1}
\begin{split}
  \frac{\Delta^{1/2}}{\sigma(x,x')^{\frac{d}{2}+1-c\epsilon}} \sim  \frac{(2\pi)^{\frac{d}{2}}\mu^{-2c\epsilon}}{c ~\epsilon ~ d ~ \Gamma(d/2)} ~\left(\nabla^2 - \frac{R}{6}\right)\delta^{(d)}(x,x')\,,
\end{split}
\end{equation}
which is obtained inverting \eqref{eq:powers-of-sigma-relation} for $b=d/2$ and using the coincidence limits of the biscalars $[\Delta^{1/2}]=1$ and $[Y]=R/6$.

Generalizations of \eqref{eq:basic-divergence+1} including higher inverse powers can also be easily obtained,
however further iterations of \eqref{eq:powers-of-sigma-relation} typically exhibit bilocal operators which are separated
from the Dirac delta by the presence of covariant derivatives (imagine, for example, placing $Q(x,x')$ on both sides of \eqref{eq:basic-divergence+1})
and hence their coincidence limit cannot be taken. In such cases, it is necessary to integrate by parts all covariant derivatives
so that all bilocal operators come in contact with the Dirac delta.
For example, if one covariant derivative is located between the bilocal operator and the Delta
we manipulate as follows
\begin{equation}\label{eq:integration-by-parts}
\begin{split}
  &Q(x,x')\nabla_\mu \delta^{(d)}(x,x') \\
  &\quad = \nabla_\mu\left(Q(x,x') \delta^{(d)}(x,x')\right) - \nabla_\mu Q(x,x') \delta^{(d)}(x,x') \\
  &\quad \sim \nabla_\mu\left([Q]\delta^{(d)}(x,x')\right) - [\nabla_\mu Q] \delta^{(d)}(x,x')\,.
\end{split}
\end{equation}
In the second line we have exploited the Delta to take the coincidence limit of the neighboring operators.
A similar manipulation can be performed for the case of two derivatives and results in
\begin{equation}
\begin{split}
  &Q(x,x')\nabla_\mu \nabla_\nu\delta^{(d)}(x,x') \\
  &\quad \sim
  \nabla_\mu \nabla_\nu \left([Q]\delta^{(d)}(x,x')\right)
  + [\nabla_\nu \nabla_\mu Q]\delta^{(d)}(x,x')
  \\
  &\quad -\nabla_\nu \left( [\nabla_\mu Q] \delta^{(d)}(x,x')\right) - \nabla_\mu \left( [\nabla_\mu Q] \delta^{(d)}(x,x')\right)
  \,.
\end{split}
\end{equation}
In order to obtain further generalizations, one has to integrate by parts all derivatives one-by-one,
and take the coincidence limits only of operators which are in direct contact with the Dirac delta.
Generalizations of \eqref{eq:integration-by-parts} are thus straightforward but rather lenghty.

Systematic applications of \eqref{eq:powers-of-sigma-relation}, to manipulate the inverse powers of the world function,
and of \eqref{eq:integration-by-parts}, to take the local parts of the biscalars multiplying the divergences,
can reduce the divergence part of the arbitrary expression \eqref{eq:arbitrary-diagram} into a simple sum of dimensionally regulated poles.

We illustrate the use of the formulas derived in the appendix for the process of dimensional regularization
showing the basics steps involved in explicitly isolating the diverging part of the first diagram in \eqref{eq:counterterms}.
We recall that the leading order
renormalization comes from $n$ propagators and that the upper critical dimension of the model $\phi^{2n}$ is $d_n=2n/(n-1)$.
The diagram is thus given by the
$n$th power of the leading term of the covariant Green function \eqref{eq:green-sdw-expansion}.
The integrand is proportional to
\begin{equation}
	\begin{split}
 		\frac{1}{\sigma(x, x')^{n\left(\frac{d}{2}-1\right)}} = \frac{1}{\sigma(x, x')^{n\frac{d_n}{2}-n-\frac{n}{2}\epsilon}}\,.
	\end{split}
\end{equation}
Our task is to cast the inverse power of the Synge function on the right hand side to match either $\frac{d}{2}$ or any integer displacement of the latter.
Using again $d=d_n-\epsilon$ and the explicit form of the upper critical dimension we find
\begin{equation}
	\begin{split}
		& n\frac{d_n}{2}-n-\frac{n}{2}\epsilon  = \frac{n^2-n^2+n}{n-1}-\frac{n}{2}\epsilon\\
		& \quad =\frac{d_n}{2}-\frac{n}{2}\epsilon = \frac{d}{2}-\frac{n-1}{2}\epsilon\,.
	\end{split}
\end{equation}
As anticipated we could identify the leading part of the exponent to be $\frac{d}{2}$, which allows us to use \eqref{eq:basic-divergence}.
This is not a coincidence as it is related to the superficial degree of divergence of the diagram under consideration;
in practice we find a pole because $n\Bigl(\frac{d}{2}-1\Bigr) \sim \frac{d}{2}$ for $\epsilon\sim0$.
We are finally lead to
\begin{equation}\label{eq:sigma-asymptotic-melonic}
	\begin{split}
		\frac{1}{\sigma(x, x')^{n\frac{d_n}{2}-n-\frac{n}{2}\epsilon}} &= \frac{1}{\sigma(x, x')^{\frac{d}{2}-\frac{n-1}{2}\epsilon}}
		\\
		&\hspace{-1cm} \underset{\epsilon\to 0}{\sim} \frac{(2\pi)^{\frac{d}{2}}}{(n-1)\,\epsilon\,\Gamma\left(d/2\right)}\mu^{(1-n)\epsilon}\delta^{(d)}(x, x')\,,
	\end{split}
\end{equation}
which was used to evaluate the right hand side of \eqref{eq:counterterms}.
Similar steps can be followed to evaluate the other two diagrams of \eqref{eq:counterterms} which exhibit the pole given by \eqref{eq:basic-divergence+1}
because their leading power is $\frac{d}{2}+1$.


\end{document}